# Noise-resolution uncertainty principle in classical and quantum systems

Timur E. Gureyev,[1,2,3,4,5] Alexander Kozlov,[1] Yakov I. Nesterets,[4,3] David M. Paganin,[2] and Harry M. Quiney[1]

[1] *ARC Centre in Advanced Molecular Imaging, School of Physics, The University of Melbourne, Parkville 3010, Australia*

[2] *School of Physics and Astronomy, Monash University, Clayton 3800, Australia*

[3] *School of Science and Technology, University of New England, Armidale 2351, Australia*

[4] *Commonwealth Scientific and Industrial Research Organisation, Clayton 3168, Australia*

[5] *Faculty of Health Science, The University of Sydney, Sydney 2006, Australia*

**Abstract:** It is proved that the width of a function and the width of the distribution of its values cannot be made arbitrarily small simultaneously. In the case of ergodic stochastic processes, an ensuing uncertainty relationship is demonstrated for the product of correlation length and variance. A closely related uncertainty principle is also established for the average degree of fourth-order coherence and the spatial width of modes of bosonic quantum fields. However, it is shown that, in the case of stochastic and quantum observables, certain non-classical states with sub-Poissonian statistics, such as for example photon number squeezed states in quantum optics, can overcome the "classical" noise-resolution uncertainty limit. The described uncertainty relationship, which is fundamentally different from the Heisenberg and related uncertainty principles, can define an upper limit for the information capacity of communication and imaging systems. It is expected to be useful in a variety of problems in classical and quantum physics.

## 1. Introduction

Various forms of uncertainty principle play important role in different branches of physics. Well-known examples are the classical diffraction limit in optics [1] and the Heisenberg uncertainty principle (HUP) in quantum physics [2]. These two principles are related to the same mathematical result [3], as are several other uncertainty relationships in physics. Due to these common foundations, the corresponding "phase volumes" are minimized by minimum uncertainty states with similar mathematical form related, for example, to Gaussian or Poisson distributions. Remarkably, a fundamentally different inequality, termed "noise-resolution uncertainty" (NRU), has very recently been discovered [4,5]. In this context, the "signal" is the mean value of a measured quantity, with the "noise" being the root-mean-square deviation from the mean. The "resolution" can be associated, for example, with the width of the point-spread function of a measuring system or with the width of a bosonic field mode. The NRU inequality is minimized by a type of state related to the Epanechnikov rather than Gaussian distributions [6,4]. This provides a strong indication that, unlike number-phase and many other uncertainty relationships in physics, the NRU (which formally corresponds to number-position uncertainty) cannot be derived from the HUP or similar relationships.

In its most general form, the NRU states that the width of a function and the width of the distribution of its values cannot be simultaneously made arbitrarily narrow. When applied to stochastic distributions, the NRU implies that a distribution, such as a detected spatial distribution of identical particles, cannot be made arbitrarily spatially narrow and have arbitrary high signal-to-noise ratio (SNR) at the same time, if the total mean number of particles in the system is kept constant. This implies a fundamental trade-off between the (spatial) resolution and the SNR of a distributed classical or quantum measurement. This may seem counter-intuitive at first in view of the obvious existence of Dirac-delta-like distributions, which may be perceived as having both "perfect" localization and arbitrarily high SNR. Nevertheless, the applicability of the NRU principle to such distributions is demonstrated below.

We show here that the NRU principle can be expressed in a general form which is different from and has broader applicability compared to our earlier results [4,5,7]. Related results have been reported previously in the field of classical imaging [8-10], quantum optics [11-13] and quantum metrology [14]. The existence of a noise-resolution trade-off has been long recognised [15], and some related notions such as, for example, the Detective Quantum Efficiency (DQE) in detector technology, have been introduced and successfully used in practice [16]. The existence and precise quantitative form of NRU-related trade-offs can be important in many fields of science and technology, from medical imaging and semiconductor manufacturing to quantum computing and experiments with ultra-cold atoms. It is both interesting and non-trivial to establish a general and precise expression for the NRU in a transparent form that can be readily related to typical settings of many different physical measurements. This question is addressed below. Following a general mathematical description, we demonstrate how the proposed approach can be applied to stochastic processes, where it provides a relationship between correlation length and variance. As another application, we derive NRU relationships for the correlation functions in bosonic quantum fields. We then give examples of non-classical quantum systems that can overcome the classical limit for the product of the variance and the width of a mode.

## 2. Relationship between the widths of a function and its histogram

Consider an arbitrary measurable (in the sense of Radon) function $f(x) \geq 0$ of real variable $x$, such that $f(x) = 0$ outside some finite interval $\Omega = [a,b]$. We study the one-dimensional case first for simplicity, but use notation, such as $|\Omega| \equiv b - a$ for the length of $\Omega$, that allows easy translation of the main results to higher dimensions later. The (normalized) histogram of $f(x)$ can be defined in a similar way to the probability density function (PDF). Firstly, consider a function $L_f(y)$ equal to the fraction (relative area) of all points $x$ in $\Omega$ at which $f(x) \leq y$, i.e. $L_f(y) = |\Omega|^{-1} \int_\Omega \theta[y - f(x)] dx$, where $\theta(x)$ is the Heaviside step-function which is equal to zero, when $x < 0$, and equal to 1, when $x \geq 0$. The normalized histogram function is then defined as the (generalized) derivative of $L_f(y)$, that is $\lambda_f(y) \equiv dL_f(y)/dy = |\Omega|^{-1} \int_\Omega \delta[y - f(x)]dx$, where $\delta(x)$ is the Dirac delta (Fig. 1). As $L_f(y)$ is a monotonically non-decreasing function of $y$, $\lambda_f(y)$ is non-negative everywhere and $\int \lambda_f(y)dy = L_f(+\infty) - L_f(-\infty) = 1$. Therefore, $\lambda_f(y)$ can be viewed as a PDF associated with the "random variable" $f(x)$. Such an interpretation can be literal e.g. if $f(x)$ is a sample of a spatially-ergodic stochastic process $\{f(x)\}$.

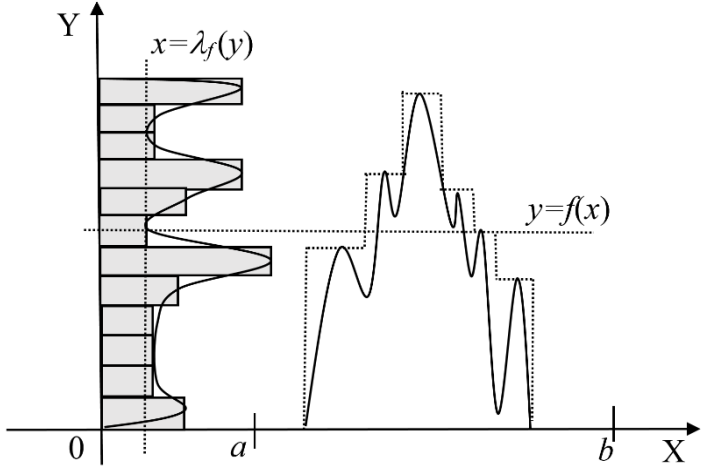

Fig.1. A graph of a function and its histogram.

A useful result involving the histogram of a function is the following equation for mean values:

$$|\Omega|^{-1} \int_\Omega g[f(x)]dx = \int g(y)\lambda_f(y)dy, \qquad (1)$$

where $g(y)$ is an arbitrary function selected from a suitably broad functional class. For example, when $g(y) = y$, we have $\overline{f} \equiv |\Omega|^{-1} \int_\Omega f(x)dx = \int y\lambda_f(y)dy$. Therefore, Eq. (1) is sometimes called "the law of the unconscious statistician" [17], because it states an intuitive fact that a mean value of a function can be equally evaluated either by averaging its "trial" values, or by integrating over the relevant PDF. In the present setting, Eq. (1) is easily proved by substituting $\lambda_f(y) = |\Omega|^{-1} \int_\Omega \delta[f(x) - y]dx$ on the right-hand side of Eq. (1) and changing the order of integration.

We define the width of a function $f(x) \geq 0$ in the usual way as

$$\Delta_x[f] \equiv \left[ \int_\Omega (x-\bar{x})^2 f(x) dx / \int_\Omega f(x) dx \right]^{1/2}, \qquad (2)$$

where $\bar{x} \equiv \int_\Omega x f(x) dx / \int_\Omega f(x) dx$. We would like to quantify and prove the following intuitively obvious hypothesis: when $f(x)$ becomes narrow, its histogram $\lambda_f(y)$ becomes broad, and vice versa. In other words, it is impossible to make a function and its histogram arbitrarily narrow, simultaneously. In accordance with this hypothesis, one could try to show that the product of the widths of $f(x)$ and $\lambda_f(y)$, $\Delta_x[f]\Delta_y[\lambda_f]$, is always larger than some positive value. Unfortunately, this is clearly impossible for finite intervals $\Omega = [a,b]$, because the width of the histogram is zero for constant functions $f(x) = c$, while the width of such functions is finite. However, it turns out that the following closely related inequality does indeed hold for any non-negative function $f$, for which the involved integrals are well defined:

$$\Delta_x[f](\Delta_y^2[\lambda_f] + \bar{f}^2) \geq C_1' |\Omega| \bar{f}^2, \qquad (3)$$

where $C_1' \cong 0.27$ is a dimensionless constant which is precisely defined below. Note that Eq. (3) holds for constant functions $f(x) = c$, because $\Delta_x[c]/|\Omega| = 1/(2\sqrt{3}) \cong 0.29 > C_1'$ for any constant $c > 0$. Moreover, Eq. (3) is exact, i.e. it becomes an equality for certain non-negative functions, such as the Epanechnikov function $f_E(x) = 1 - (2x - a - b)^2 / (b-a)^2$. A general proof of Eq. (3) follows from the identity $\Delta_y^2[\lambda_f] + \bar{f}^2 = \int y^2 \lambda_f(y) dy = |\Omega|^{-1} \int_\Omega f^2(x) dx$ that can be obtained from Eq. (1) with $g(y) = y^2$, and the one-dimensional ($d = 1$) version of the mathematical inequality:

$$\| f \|_2^2 \, \Delta_x^d[f] \geq C_d' \, \| f \|_1^2, \qquad (4)$$

where $\| f \|_p \equiv (\int |f(x)|^p \, dx)^{1/p}$ for any $p > 0$, $C_d' \equiv [d/(4\pi)]^{d/2} C_d$ and $C_d = 2^d \Gamma(d/2) d(d+2)/(d+4)^{d/2+1}$ is the Epanechnikov constant [5]. The constants $C_d$ are of the order of 1 for low dimensions $d$; in particular, $C_1 = (6/5)\sqrt{\pi/5} \cong 0.95$ and $C_1' = C_1/(2\sqrt{\pi}) = 3/(5\sqrt{5}) \cong 0.27$. Equation (4) is invariant with respect to multiplicative scaling of $f(x)$ and its argument, i.e. Eq. (4) does not change if we replace $f(x)$ with $\alpha f(\beta x)$, where $\alpha$ and $\beta$ are arbitrary real positive constants. This scaling bi-invariance of Eq. (4) is the same as that of the HUP [4].

Consider now the "variance" of a function defined in the conventional way:

$$\text{Var}[f] \equiv |\Omega|^{-1} \int_\Omega (f - \overline{f})^2(x)dx = \int (y - \overline{y})^2 \lambda_f(y)dy, \qquad (5)$$

where we have used Eq. (1) with $g(y) = (y - \overline{y})^2$. Equation (5) implies, in particular, that the variance of a function is equal to the square of the width of its histogram: $\text{Var}[f] = \Delta_y^2[\lambda_f]$. Hence, eq. (3) can be re-written as

$$\left( \frac{\text{Var}[f]}{(\overline{f})^2} + 1 \right) \frac{\Delta_x[f]}{|\Omega|} \geq C_1'. \qquad (6)$$

The first term in Eq. (6) can be identified with the squared "noise-to-signal ratio" (NSR), $\text{NSR}^2 \equiv \text{Var}[f]/(\overline{f})^2$, i.e. the inverse of the squared SNR, with $\text{SNR}^2$ defined as the ratio of the squared mean of a function to its variance. The term outside the brackets on the left-hand side of Eq. (6) is equal to $1/M$, with $M = |\Omega|/\Delta_x[f]$ being the number of widths of $f(x)$ that fit into $\Omega$. Using this intuitive and dimensionless notation, Eq. (6) can be re-stated as

$$M\,\text{SNR}^2 / (\text{SNR}^2 + 1) \leq 1/C_1'. \qquad (7)$$

If one tries to make the SNR large (which is often desired in experiments), then $\text{SNR}^2/(\text{SNR}^2+1) \cong 1$, and Eq. (7) implies that $f(x)$ must be broad, in the sense that $M \leq (\text{NSR}^2 + 1)/C_1' \cong 4$, since $\text{NSR}^2 \ll 1$. Alternatively, if one tries to squeeze $f(x)$ to achieve high spatial localization, making $M \gg 1$ as a result, then Eq. (7) implies that $\text{NSR}^2 + 1 \geq MC_1' \gg 1$ and hence $\text{SNR}^2 \ll 1$. Therefore, a trade-off always exists between the SNR and spatial resolution and both of them cannot be improved simultaneously beyond a certain limit. This relationship is a modified and generalized version of the previously proposed NRU principle [4,5]. The lower limit of the left-hand side of Eq. (6), which is equal exactly to $C_1'$, is achieved for suitable Epanechnikov distributions such as $f_E(x)$ defined above. The above results can be easily extended to arbitrary dimensions $d$ [5]. In particular, Eq. (7) becomes $M\,\text{SNR}^2 / (\text{SNR}^2 + 1) \leq 1/C_d'$.

It is instructive to consider the NRU Eq. (6) for Dirac-delta type functions. If $f(x) = \delta(x)$, then $\text{Var}[f]$ cannot be formally evaluated, because the square of the delta function is not defined in the space of generalized functions. Consider, however, an area $\Omega_b \equiv [-b,b]$ and the set of Gaussians $f_\sigma(x) = (2\pi)^{-1/2} \sigma^{-1} \exp[-|x|^2/(2\sigma^2)]$ with $\sigma \ll b$, which approximate $\delta(x)$ when $\sigma \to 0$. It is easy to calculate that $\overline{f}_\sigma \cong 1/(2b)$, $\text{Var}[f_\sigma] \cong 1/(4b\sigma\sqrt{\pi})$, $\Delta_x[f_\sigma] \cong \sigma$, and hence in the left-hand side of Eq. (6) we have $\{1+\text{Var}[f_\sigma]/(\overline{f}_\sigma)^2\}\Delta_x[f_\sigma]/(2b) \cong (2\sqrt{\pi})^{-1} + \sigma/(2b)$. This value tends to the constant $(2\sqrt{\pi})^{-1} \cong 0.28 > C_1'$, when $\sigma \to 0$, and, in particular, the NRU does not decrease below this positive value when the Gaussians approach the Dirac delta function.

## 3. Noise-resolution uncertainty for stochastic processes

Let us consider an application of the above result to ergodic stochastic distributions (spatial stochastic processes). Here it is convenient to work with symmetric domains $\Omega_b = [-b, b]$. To keep the notation consistent with that used above, instead of considering the proper limits of various quantities at $b \to \infty$ as required in the theory of stochastic processes, we simply assume that $b$ is sufficiently large. For example, the autocorrelation function will be defined as $\Gamma_f(x) = |\Omega_b|^{-1} \int_{\Omega_b} f(x+x')f(x')dx'$, implicitly assuming that $b$ is so large that contribution from the "tails" of the integral at $|x| \geq b$ is negligible for all considered distributions $f(x)$.

Because ergodic distributions $f(x)$ are always stationary [18], the notion of width does not make sense for them. However, the width of $\Gamma_f(x)$, i.e. the correlation length [18], is well defined. We consider here a frequently encountered case, where $f(x)$ is a convolution of an "input" distribution $f_{in}(x)$ with a deterministic point-spread function $P(x)$, $f(x) = (f_{in} * P)(x) = \int f_{in}(x-x')P(x')dx'$. For each distribution, we will also consider the corresponding "noise" distribution $\tilde{f}(x) = f(x) - \overline{f}$. When the noise $\tilde{f}_{in}(x)$ in the input distribution is essentially uncorrelated, its correlation length is very small and the correlation length of $\tilde{f} = \tilde{f}_{in} * P$ is determined by the width of $P(x)$. Under these conditions, we will obtain an analog of the NRU for the stochastic distribution $f = f_{in} * P$ by applying Eq. (4) to $P(x)$.

Firstly, note that $\overline{f} = |\Omega_b|^{-1} \int_{\Omega_b} \int f_{in}(x-x')P(x')dx'dx \cong \overline{f}_{in} \|P\|_1$. It is also easy to verify that $\Gamma_{\tilde{f}}(x) = (\Gamma_{\tilde{f}_{in}} * P_2)(x)$, where $P_2(x) \equiv \int P(x+x')P(x')dx'$ is the autocorrelation of $P(x)$. Therefore, $\text{Var}[f] = \Gamma_{\tilde{f}}(0) = \int \Gamma_{\tilde{f}_{in}}(-x')P_2(x')dx' \cong \Gamma_{\tilde{f}_{in}}(0)h_{in}P_2(0) = \text{Var}[f_{in}]h_{in}\|P\|_2^2$, where we have used the fact that $\tilde{f}_{in}(x)$ is uncorrelated and hence $\Gamma_{\tilde{f}_{in}}(y)$ is very narrow (we denoted its width by $h_{in}$). Combining the above results, we obtain $\text{Var}[f]/(\overline{f})^2 = h_{in}\|P\|_2^2/(\|P\|_1^2 \text{SNR}_{in}^2)$, where $\text{SNR}_{in} \equiv \overline{f}_{in}/\sigma_{in}$. Multiplying this equation

by $\Delta_x[P]$ and applying Eq. (4) with $d=1$ for $P(x)$, $\|P\|_2^2 \Delta_x[P] \geq C_1' \|P\|_1^2$, leads to the following analog of Eq. (6):

$$\frac{\text{Var}[f]}{(\bar{f})^2} \frac{\Delta_x[P]}{|\Omega_b|} \geq \frac{C_1'}{M_{in} \text{SNR}_{in}^2}, \qquad (8)$$

where $M_{in} \equiv |\Omega_b|/h_{in}$ is the number of autocorrelation lengths of the input distribution that fit into $\Omega_b$. Consequently, instead of Eq. (7), we here obtain:

$$M_P \text{SNR}^2 \leq (C_1')^{-1} M_{in} \text{SNR}_{in}^2, \qquad (9)$$

where $M_P \equiv |\Omega_b|/\Delta_x[P]$. The difference in form between Eq. (6) and Eq. (8) includes the absence of the additive term 1 inside the brackets on the left-hand side and the presence of the term $M_{in} \text{SNR}_{in}^2$ on the right-hand side. If $f_{in}$ has large SNR$_{in}$ and/or large $M_{in}$, then the right-hand side of Eq. (8) can be close to zero; in which case Eq. (8) does not impose any substantial lower limits on the product of the NSR$^2$ and the width of $P(x)$. We see below that a similar situation can occur for quantum fields.

In relation to imaging problems, note that the structure of Eq. (9) is reminiscent of the Detective Quantum Efficiency (DQE) at zero frequency, which describes the efficiency of a detector and is usually equal to the ratio of the squared SNR in output and input signals [16]. A special case of Eq. (9) corresponds to $f_{in}(x)$ being an uncorrelated distribution with mean number of photons $\bar{n}_{in}$ collected in each detector pixel and $\text{SNR}_{in}^2 = \bar{n}_{in}^\gamma$, where $0 \leq \gamma < 1$ in the case of super-Poissonian statistics ($\gamma = 0$ for Gaussian statistics), $\gamma = 1$ for Poissonian statistics and $1 < \gamma < \infty$ for sub-Poissonian statistics [19]. Here Eq. (9) becomes [7]:

$$\frac{\text{SNR}^2}{\Delta_x[P]} \leq \frac{\bar{n}_{in}^\gamma}{C_1' h_{in}}. \qquad (10)$$

Equation (10) with $\gamma = 1$ expresses a typical form of NRU in imaging, where, at a fixed incident photon fluence $F_{in} = \bar{n}_{in}/h_{in}$, the ratio of the squared SNR to spatial resolution cannot be increased beyond a certain absolute limit. More generally, if $\bar{n}_{in} > 1$ and $\gamma \gg 1$, $\bar{n}_{in}^\gamma$ can in principle be very large and the SNR can be essentially decoupled from the correlation length of a stochastic distribution, at a fixed incident fluence (radiation dose) level. A similar phenomenon is also observed in the context of quantum field theory, as demonstrated below. This can be advantageous in imaging of radiation-sensitive samples [14].

We note that the above results have direct implications for the information capacity of imaging systems [20] and possibly for quantum information capacity as well [21]. In particular, Eq. (9) provides an upper bound for the Shannon information capacity $C_S$ of an imaging system with $M_P$ independent channels (effective pixels) and Poissonian noise statistics, $\text{SNR}_{in}^2 = \bar{n}_{in}$:

$$C_S \cong 0.5 M_P \log_2(SNR^2 + 1) \leq 0.5 M_P SNR^2 \leq \bar{N}_{in} / (2C_1'), \tag{11}$$

where $\bar{N}_{in} \equiv M_{in} \bar{n}_{in}$ denotes the mean total number of photons used for imaging. Equation (11) implies that the Shannon information capacity per photon is limited from above by an absolute constant $(2C_1')^{-1} \cong 1.9$.

## 4. Noise-resolution uncertainty in quantum field theory

Let us now consider an analog of the NRU Eq. (6) in quantum electrodynamics (QED). For a single-mode electric field, $E_k(\mathbf{r},t) = E_k^{(+)}(\mathbf{r},t) + E_k^{(-)}(\mathbf{r},t)$ inside a cube $\Omega_L$ with side length $L$ centered at $\mathbf{r} = 0$ in 3D space, $E_k^{(+)}(\mathbf{r},t) = i(\hbar \omega_k / 2)^{1/2} u_k(\mathbf{r}) \exp(-i\omega_k t) \hat{a}_k$, where $\hbar$ is the reduced Planck constant, $\omega_k$ is the angular frequency of the mode, $u_k(\mathbf{r})$ is the mode function, $E^{(-)}(\mathbf{r},t) = [E^{(+)}(\mathbf{r},t)]^\dagger$ is the Hermitian conjugate of $E^{(+)}(\mathbf{r},t)$, $\hat{a}_k$ and $\hat{a}_k^\dagger$ are the photon annihilation and creation operators for the $k$th mode, respectively [19].

If $\rho$ is a density operator and $<n_k>$ is the mean number of photons in the mode, then $\text{Tr}[\rho \hat{a}_k^\dagger \hat{a}_k] = <n_k>$ and $\text{Tr}[\rho \hat{a}_k^\dagger \hat{a}_k^\dagger \hat{a}_k \hat{a}_k] = \text{Tr}[\rho (\hat{a}_k^\dagger \hat{a}_k)^2] - \text{Tr}[\rho \hat{a}_k^\dagger \hat{a}_k] = <n_k^2> - <n_k>$, where we have used the commutation relation $[\hat{a}_k, \hat{a}_k^\dagger] = 1$ [19]. Therefore, Glauber's coherence functions of the second and fourth order [22,19], in the special case of single-point measurements, can be expressed as

$$G_1(\mathbf{r}) = \text{Tr}[\rho E_k^{(-)}(\mathbf{r},t) E_k^{(+)}(\mathbf{r},t)] = <n_k> (\hbar \omega_k / 2) f_k(\mathbf{r}), \tag{12}$$

$$G_2(\mathbf{r}) \equiv \mathrm{Tr}[\rho E_k^{(-)}(\mathbf{r},t) E_k^{(-)}(\mathbf{r},t) E_k^{(+)}(\mathbf{r},t) E_k^{(+)}(\mathbf{r},t)]$$
$$= (<n_k^2> - <n_k>)(\hbar\omega_k/2)^2 f_k^2(\mathbf{r}), \quad (13)$$

where $f_k(\mathbf{r}) \equiv |u_k(\mathbf{r})|^2$. For any function $G(\mathbf{r})$ defined in the three-dimensional area $\Omega_L$ we can consider the spatial average $\overline{G} \equiv |\Omega_L|^{-1} \int_{\Omega_L} G(\mathbf{r})d\mathbf{r}$, $|\Omega_L| = L^3$, and the width $\Delta_r[G] = [\int_{\Omega_L} |\mathbf{r}-\overline{\mathbf{r}}|^2 G(\mathbf{r})d\mathbf{r} / \int_{\Omega_L} G(\mathbf{r})d\mathbf{r}]^{1/2}$, where $\overline{\mathbf{r}}$ is the mean value of $\mathbf{r}$ with respect to the PDF $G(\mathbf{r})/\int_{\Omega_L} G(\mathbf{r})d\mathbf{r}$. This allows us to introduce the quantities $\overline{G}_1 = <n_k>(\hbar\omega_k/2)\overline{f_k}$, $\overline{G}_2 = [<n_k^2> - <n_k>](\hbar\omega_k/2)^2 \overline{f_k^2}$, and consider the following analog of the functional on the left-hand side of the NRU Eq. (6):

$$F[E] \equiv \frac{\overline{G}_2}{(\overline{G}_1)^2} \frac{(\Delta_r[G_1])^3}{|\Omega_L|} = \left[\frac{\overline{G}_2 - (\overline{G}_1)^2}{(\overline{G}_1)^2} + 1\right] \frac{(\Delta_r[G_1])^3}{|\Omega_L|}. \quad (14)$$

It is easy to verify that

$$F[E] = \left(\frac{<n_k^2> - <n_k>}{<n_k>^2}\right) \frac{\int_{\Omega_L} f_k^2(\mathbf{r})d\mathbf{r}}{(\int_{\Omega_L} f_k(\mathbf{r})d\mathbf{r})^2} \left[\frac{\int_{\Omega_L} |\mathbf{r}-\overline{\mathbf{r}}|^2 f_k(\mathbf{r})d\mathbf{r}}{\int_{\Omega_L} f_k(\mathbf{r})d\mathbf{r}}\right]^{3/2}. \quad (15)$$

Applying Eq. (4) with $d=3$ to $f_k(\mathbf{r})$ in Eq. (15), we obtain the following form of the NRU:

$$\left[\frac{\overline{G}_2 - (\overline{G}_1)^2}{(\overline{G}_1)^2} + 1\right] \frac{\Delta_r^3[G_1]}{|\Omega_L|} \geq C_3'\left(1 + \frac{Q}{<n_k>}\right), \quad (16)$$

where $C_3' = [3/(4\pi)]^{3/2} C_3 = (3/7)^{1/2} 45/(98\pi) \cong 0.1$ and $Q \equiv (<(\Delta n_k)^2>/<n_k>) - 1$ is the Mandel $Q$ parameter [23,24]. For Poissonian statistics, e.g. for coherent states, $Q = 0$ and Eq.

(16) transforms back into the "classical" form of the NRU. In non-classical quantum states with sub-Poissonian statistics $Q$ becomes negative, and Eq. (16) implies that it may be possible, in principle, to make a mode arbitrarily narrow (i.e. achieve arbitrarily fine spatial resolution in a corresponding experiment), and at the same time have arbitrary small "average variance", $\overline{G}_2 - (\overline{G}_1)^2$.

Consider the specific values that the NRU, Eq. (16), implies for the lower limit of the product of the average degree of fourth-order coherence, $\overline{G}_2 / (\overline{G}_1)^2$, and the width of the mode, in an experiment where measurements are performed by a two-dimensional position-sensitive detector located near the origin of coordinates in the plane $(\mathbf{r}_\perp, z=0)$. Let a mode have a Gaussian spatial distribution in the detector plane, $f_k(\mathbf{r}_\perp, 0) = (2\pi\sigma^2)^{-1} \exp[-|\mathbf{r}_\perp|^2/(2\sigma^2)]$. The corresponding two-dimensional version of Eq. (16) similarly follows from Eq. (4) with $d = 2$. Compared to Eq. (16), we need only replace the relative volume with the relative area, $\Delta_r^2[G_1(z=0)]/L^2$, and the constant $C_3'$ with $C_2' = 4/(9\pi) \cong 0.14$. Now consider the two-dimensional version of Eq. (15) for Poissonian statistics, $(<n_k^2> - <n_k>)/<n_k>^2 = 1$, and a Gaussian-shape mode. As the functional $F[E]$ is bi-invariant with respect to the scaling of the mode function and its argument, its value for a Gaussian mode is independent of $\sigma$. If $\sigma << L$, evaluation of the relevant integrals leads to: $\overline{G}_2 / (\overline{G}_1)^2 \cong 1/(4\pi\sigma^2)$, $\Delta_r^2[G_1(z=0)]/L^2 \cong 2\sigma^2$ and hence $F[E] = 1/(2\pi) \cong 0.16 > C_2'$, regardless of $\sigma$. For an Epanechnikov-shape mode, $f_k(\mathbf{r}_\perp, 0) = (1 - 4|\mathbf{r}_\perp|^2/L^2)_+$, where the subscript "+" denotes that $f_k(\mathbf{r}_\perp, 0) = 0$ when $|\mathbf{r}_\perp| > L/2$, and Poissonian statistics of the photon state, one obtains the minimal possible value of $F[E] = C_2'$.

Consider the photon number squeezed states $|\alpha, \theta> \equiv \hat{D}(\alpha)\hat{S}(\theta)|0>$, where $\hat{D}(\alpha) \equiv \exp(\alpha\hat{a}^\dagger - \alpha^*\hat{a})$ is the unitary displacement operator, $\hat{S}(\theta) \equiv \exp[(1/2)(\theta^*\hat{a}^2 - \theta\hat{a}^{\dagger 2})]$ is the squeeze operator, $\alpha, \theta$ are complex numbers, $|0>$ is the Fock vacuum, $|\theta|$ is the squeeze parameter and $<n_k> = |\alpha|^2 + |\sinh|\theta||^2$ [19]. It can be shown that when $|\alpha|^2 >> \exp(6|\theta|)$, the Mandel parameter can be negative: $Q = Q(\alpha, \theta) < [-1 + \exp(-2|\theta|)] < 0$. In this case, the factor, $1 + Q/<n_k>$, multiplying the constant $C_3'$ on the right-hand side of Eq. (16), is less than one, meaning that the "classical" lower limit of $C_3'$ can be overcome, albeit only by a small margin.

Finally, we note that, using our earlier results [7], it can be shown that the NRU for the spatially-averaged variance of the electric energy operator, $\text{Var}[E^2] \equiv \overline{<E^4>} - (\overline{<E^2>})^2$ can be written in a form similar to Eq. (16), but with the right-hand side always limited from below by an absolute positive constant determined by vacuum fluctuations.

## 5. Conclusions

We have demonstrated that the width of a function and the width of its histogram cannot be made arbitrarily small at the same time. This relationship can be also stated in terms of an uncertainty relationship between the spatial resolution and the signal-to-noise ratio of a distributed measurement. In the case of statistical quantities associated with stochastic or quantum observables, for example, correlation functions of bosonic fields, the NRU can in principle be made arbitrarily small for non-classical states with sub-Poissonian statistics. We have shown that photon number squeezed states in quantum optics present an example where the classical limit for the NRU can be overcome.